\documentclass[onecolumn,authoryear]{els-mrw} 

\usepackage{amsmath,amssymb,amsfonts,amsthm,makeidx,graphicx}
\usepackage{txfonts}
\usepackage{helvet}
\usepackage{url}

\makeatletter
\renewcommand\tableofcontents{%
    \section*{\contentsname
        \@mkboth{%
           \MakeUppercase\contentsname}{\MakeUppercase\contentsname}}%
    \@starttoc{toc}%
    }
\makeatother

\begin{document}

\chapter*{White dwarf fundamentals}

\author[1]{Simon Blouin\footnote{sblouin@uvic.ca}}

\address[1]{\orgname{University of Victoria}, \orgdiv{Department of Physics and Astronomy}, \orgaddress{Victoria, BC V8W 2Y2, Canada}}

\maketitle

\begin{abstract}[Abstract]
This chapter provides an in-depth overview of white dwarfs, the evolutionary terminus of the vast majority of stars. It discusses their discovery, their nature as degenerate objects, their connections to earlier phases of stellar evolution, their subsequent evolution as they gradually cool down, the varied physical conditions from their dense cores to their tenuous atmospheres, some key statistics about the properties of the ever expanding population of known white dwarfs, the diversity of their spectra, the accretion of planetary material, and the presence of magnetic fields. The chapter also highlights the instrumental role of white dwarfs in other areas of astronomy.
\end{abstract}

\begin{keywords}
Chandrasekhar limit, Compact objects, Degenerate matter, Stellar atmospheres, Stellar evolution, White dwarf stars
\end{keywords}


\begin{glossary}[Glossary]
\term{Chandrasekhar limit} Maximum mass that a stable white dwarf star can have before collapsing

\term{Cooling track} Theoretical model describing how a white dwarf of a particular mass and composition cools over time

\term{Core crystallization} Liquid--solid phase transition undergone by the dense plasma in white dwarf cores

\term{Degenerate matter} State of matter in which quantum mechanical effects dominate over thermal effects due to Pauli's exclusion principle

\term{Equation of state} Model describing the relationship between properties such as pressure, volume, and temperature for a given substance

\term{Gravitational settling} Process where heavier elements sink while lighter elements rise to the surface

\term{Luminosity function} Distribution showing the number of white dwarfs at different luminosities

\term{Mass--radius relation} Relation describing how higher-mass white dwarfs have smaller radii

\term{Spectral evolution} Changes in the surface composition of white dwarfs as they cool

\term{White dwarf cosmochronology} The study of cooling white dwarfs to determine the age of stellar populations

\end{glossary}

\begin{glossary}[Nomenclature]
\begin{tabular}{@{}lp{34pc}@{}}
DA & White dwarf displaying hydrogen absorption lines\\
DB & White dwarf displaying neutral helium absorption lines\\
DC & White dwarf with a featureless spectrum\\
DO & White dwarf displaying ionized helium absorption lines\\
DQ & White dwarf displaying carbon absorption features\\
DZ & White dwarfs displaying absorption lines from metals other than carbon
\end{tabular}
\end{glossary}

\subsection*{Learning objectives}
\begin{itemize}
    \item White dwarfs were first recognized as a fundamentally different class of stars in the early 1900s due to their small radii.
    \item White dwarfs are supported by electron degeneracy pressure and obey a mass--radius relationship.
    \item White dwarfs are the final stage in the evolution of the vast majority of stars.
    \item The composition profiles of white dwarfs bear the imprint of previous evolutionary phases.
    \item The evolution of white dwarfs is characterized by a long cooling process.
    \item White dwarf masses are narrowly distributed around $\simeq 0.6\,M_{\odot}$.
    \item The faint end of the white dwarf cooling sequence provides a valuable tool for measuring the ages of stellar populations.
    \item White dwarf atmospheres vary widely in temperature and composition, resulting in a broad spectrum of spectral types.
    \item The pollution of white dwarf atmospheres by planetary debris offers insights into exoplanetary systems.
    \item White dwarfs can have strong magnetic fields, the origin of which remains debated.
\end{itemize}

\section{Introduction}\label{sec:intro}
Around 98\% of stars in our Galaxy will eventually become white dwarfs at the end of their long lives, 6\% of stars in the solar neighbourhood are currently white dwarfs, and over 350{,}000 white dwarfs have now been identified with high confidence. Yet, unless they deliberately sought them out, chances are the reader of this chapter has never seen a white dwarf in the night sky. If white dwarfs are so inconspicuous despite being so prevalent, it is because they are intrinsically faint. Only about the size of the Earth, they shine orders of magnitude less brightly than more voluminous main-sequence and giant stars with the same temperature, such that none is visible to the naked eye. White dwarfs are small, but their masses are comparable to that of the Sun, making them extremely dense objects, 100{,}000 times denser than the center of the Earth.

These stellar remnants have historically proven to be exceptionally fertile grounds for advances in physics and astronomy. In the 1920s and 1930s, understanding the very nature of white dwarfs as degenerate stars constituted one of the first applications of quantum mechanics to astrophysics (this groundbreaking work was later recognized in the 1983 Nobel Prize in Physics). Since the 1980s, the oldest white dwarfs are used to place tight constraints on the ages on various stellar populations within the Milky Way. Type Ia supernovae, powerful thermonuclear detonations that occur when a white dwarf accretes too much mass from a companion, led to the paradigm-shifting discovery of the accelerating expansion of the Universe in 1998 (also later recognized with a Nobel Prize in 2011). Today, the \textit{James Webb Space Telescope} is searching for biosignatures on planets orbiting white dwarfs, the European Space Agency's \textit{Gaia} mission is enabling us to glimpse at the exotic physics of crystallization in the dense plasmas found in the cores of white dwarfs, and in the not-so-distant future \textit{LISA} will detect gravitational waves emanating from white dwarf pairs. But white dwarfs are more than useful testbeds for the physics of extreme densities, tracers of cosmology, or laboratories for astrobiology. They are also fascinating objects in their own rights, and in large part, this is what this chapter will attempt to demonstrate.

The chapter is organized as follows. Section~\ref{sec:history} presents a brief historical account of the discovery of white dwarfs. The conundrum that arose from these early observations will naturally lead us to Section~\ref{sec:nature}, which details the nature of white dwarfs as degenerate stars and introduces the famous Chandrasekhar limit. We will then look at the origins of white dwarfs in Section~\ref{sec:evol} by detailing the stellar evolution phases that shape the composition profiles of white dwarfs. In particular, this section will expand on the numerous connections between the helium-burning phases of stellar evolution and white dwarfs. We will then explore in Section~\ref{sec:cooling} how the star evolves once it has become a white dwarf. This is a vast topic in itself, and we will cover only the basic notions required for the rest of this chapter. This section will also provide an opportunity to more closely examine the physical conditions within white dwarfs, from their dense cores to their sparse atmospheres. The subsequent section, Section~\ref{sec:demographics}, delves into the analysis of white dwarfs as a population, focusing on their characteristically narrow mass distribution, as well as the importance of the luminosity function in age-dating applications. Section~\ref{sec:spectro} will then discuss important notions pertaining to the atmospheres of white dwarfs, including spectral classification, external pollution by planetary material, and magnetic fields. Finally, a brief conclusion is given in Section~\ref{sec:conclusion}.

\section{The discovery of white dwarfs}\label{sec:history}
The earliest observation of a white dwarf is attributed to William Herschel (1738--1822), who in 1783 discovered two faint companions to the much brighter 40 Eridani A (abbreviated 40 Eri A).\footnote{This section is based in large part on \cite{holberg2009} and \cite{vanhorn2015}.} We now know that one of these companions, 40 Eri B, is a white dwarf. Although it was discovered first, 40 Eri B is not the brightest white dwarf in the night sky. This distinction goes to Sirius B. However, Sirius B is much harder to observe as it is swamped by the intense brightness of its companion Sirius A, the brightest star visible at night. In fact, the existence of Sirius B was hypothesized some two decades before it could be discovered. In 1844, Wilhelm Friedrich Bessel (1784--1846) suggested that yet unseen binary companions were responsible for irregularities in the motions of Sirius and Procyon across the sky. Sirius~B was finally discovered in 1862 and found to be 10{,}000 times fainter than Sirius A.

Since the components of a given star system are at a very similar distance from the Earth, large differences in apparent brightness imply large differences in intrinsic luminosity. The luminosity $L$ of a star is given by
\begin{equation}
    L = 4 \pi R^2 \sigma T_{\rm eff}^4,
    \label{eq:sb}
\end{equation}
where $R$ is the star's radius, $\sigma$ is the Stefan--Boltzmann constant and $T_{\rm eff}$ is the effective temperature. At the time, the large difference in $L$ between Sirius A and Sirius B (and between 40 Eri A and 40 Eri B) could a priori be attributed to differences in temperatures, because their temperatures were yet unknown. With this convenient explanation, there was apparently nothing unusual left to explain. It is only decades later that the truly peculiar nature of 40 Eri B and Sirius B would become clear.

In 1910, Henry Norris Russell, Edward Charles Pickering and Williamina Fleming discovered that 40 Eri B was of spectral class A, meaning that it is hot enough to show hydrogen spectral lines. This was totally unexpected, because it meant that it had to be hotter than 40 Eri A, a K-type star. 40 Eri B being $\sim 100$ times fainter than 40 Eri A, this realization implied that 40 Eri B was remarkably small (Equation~\ref{eq:sb}). Similar conclusions were reached for Sirius B and van Maanen's Star (an isolated star, also known as van Maanen 2) during the following decade, clearly establishing a new and puzzling class of stars. This newest addition to the stellar zoo was first coined ``white dwarfs'' by \cite{luyten1922} to highlight their small size and white color (indicative of hot temperatures). In addition, because 40 Eri B and Sirius B belong to triple and binary star systems, their masses could be determined from their orbital motions. Their substantial masses ($M \sim 1 M_{\odot}$) and small radii ($R \sim 0.01\,R_{\odot}$) implied very high densities ($\rho \sim 10^6\,{\rm g}\,{\rm cm}^{-3}$) that seemed aberrant at the time, before the development of quantum mechanics. In the words of \cite{eddington1927}:
\begin{quote}
The message of the Companion of Sirius when it was decoded ran: ``I am composed of material 3{,}000 times denser than anything you have come across; a ton of my material would be a little nugget that you could put in a matchbox.'' What reply can one make to such a message? The reply which most of us made in 1914 was---``Shut up. Don't talk nonsense.''
\end{quote}
In the next section, we will make sense of this ``nonsense'' by invoking the quantum mechanical concepts that were not yet available to the discoverers of white dwarfs at the beginning of the previous century.

\section{The degenerate nature of white dwarfs}\label{sec:nature}
\subsection{Electron degeneracy pressure}
Given their high temperatures and extreme densities, white dwarfs interiors are completely ionized plasmas, where atoms are stripped bare of their electrons. Moreover, particles are so closely packed together that the average interparticle distance is smaller than the electrons' thermal de Broglie wavelength,
\begin{equation}
    \lambda_{\rm th} = \sqrt{ \frac{2 \pi \hbar^2}{m_e k_B T} },
\end{equation}
where $\hbar$ is the reduced Planck constant, $m_e$ is the electron mass, where $k_B$ is the Boltzmann's constant, and $T$ is the temperature. This is key to understanding the nature of white dwarfs, as quantum effects arise when the de Broglie wavelengths overlap. In particular, the Pauli exclusion principle significantly alters the behavior of electrons in white dwarfs. This principle states that identical fermions (e.g., electrons) cannot simultaneously occupy the same quantum state. In the extremely dense plasmas of white dwarfs, this forces electrons to occupy higher-energy levels,  leading to the formation of ``degenerate matter''. Crucially, trying to reduce the volume occupied by these degenerate electrons would force them to occupy yet higher-energy states. This would require the application of a compression force against this ``degeneracy pressure''.

For a completely degenerate electron gas (a good approximation for the extreme densities that characterize most of a white dwarf's interior) and assuming that the density is not too high (see below), this degeneracy pressure $P$ is given by
\begin{equation}
    P = \frac{\left( 3 \pi^2 \right)^{2/3}}{5} \frac{\hbar^2}{m_e} n_e^{5/3},
    \label{eq:Pdegen}
\end{equation}
where $n_e$ is the electron number density. This $P \propto \rho^{5/3}$ equation of state was derived by \cite{fowler1926}, who was investigating the nature of matter at the mean white dwarf densities established during the previous decade from astronomical observations. This was one of the first applications of quantum mechanics to astrophysics, and it contained the key insight required to understand the structure of white dwarfs. More specifically, it is this degeneracy pressure that opposes the white dwarf's strong gravity and prevents it from collapsing under its own weight. This is to be contrasted with lower-density main sequence stars, where thermal pressure (as in the ideal gas law, $P=n k_B T$) plays this role.

\subsection{The mass--radius relationship and the Chandrasekhar limit}
By combining this equation of state (Equation~\ref{eq:Pdegen}) with two of the basic equations of stellar structure, the hydrostatic equilibrium and the mass conservation equations, the famous white dwarf mass--radius relationship can be derived. In the simple case where Equation~\eqref{eq:Pdegen} applies, it takes the form $M \propto R^{-3}$. This implies that higher-mass white dwarfs have \textit{smaller} radii. The higher the mass, the stronger the gravity and the higher $n_e$ must be to provide the degeneracy pressure required to avoid collapse. There is however a limit to this relation. Equation~\eqref{eq:Pdegen} relied on the assumption that the electrons are not relativistic. However, with increasing densities, electrons are forced into higher and higher energy states and eventually exceed the rest energy of an electron, thereby becoming relativistic particles. As the electronic velocities increase, the equation of state ``softens'' and eventually becomes $P \propto \rho^{4/3}$ in the extreme relativistic limit:
\begin{equation}
    P = \frac{\left( 3 \pi^2 \right)^{1/3}}{4} \hbar c n_e^{4/3},
\end{equation}
where $c$ is the speed of light. Note that we have again assumed that the electrons are completely degenerate. For simplicity, equations for the general case of partial degeneracy and arbitrary relativity are not provided here. The softening of the equation of state from $P \propto \rho^{5/3}$ to $P \propto \rho^{4/3}$ indicates that past a certain point, increasing the star's density provides a diminishing increase in the degeneracy pressure. Ultimately, this sets an upper limit of $\simeq 1.4\,M_{\odot}$ on the mass of a stable white dwarf. This is the famous \cite{chandrasekhar1931} limit. A white dwarf with a mass exceeding this critical threshold would collapse into a neutron star or black hole.

\begin{figure}
\centering
\includegraphics[width=0.6\textwidth]{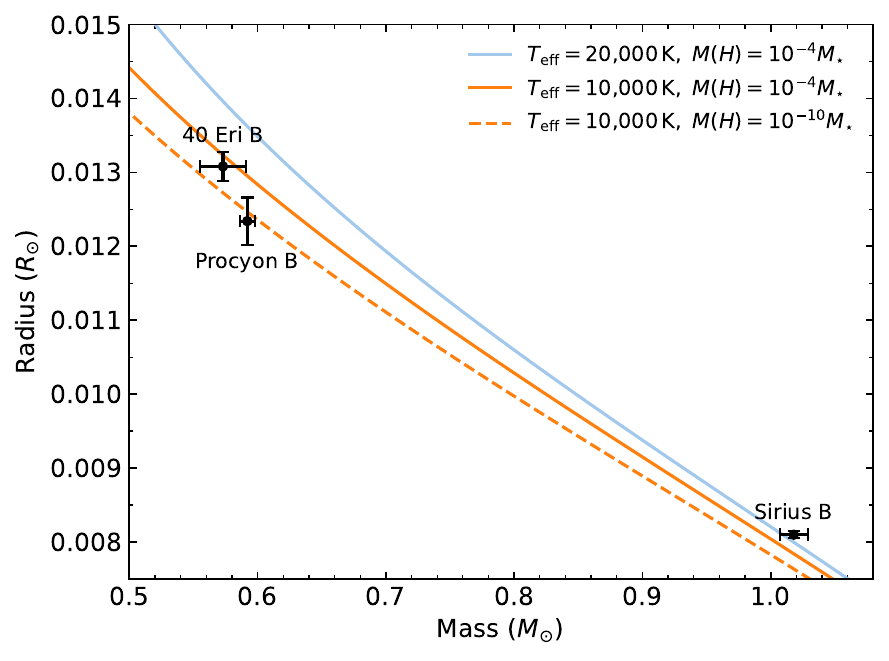}
\caption{White dwarf mass-radius relationships. The curves correspond to predictions from a modern stellar evolution code \citep{bedard2022} assuming different effective temperatures and hydrogen layer thicknesses (see legend). The masses and radii of three classical white dwarfs are also shown \citep{bond2015,bond2017a,bond2017b}.}
\label{fig:mr}
\end{figure}

Figure~\ref{fig:mr} illustrates modern theoretical mass--radius relationships. Today, detailed numerical models are used instead of the analytical relations described above. These models take into consideration the detailed composition of the white dwarf's degenerate interior, the size of its non-degenerate outer layers, and the small but non-negligible effects of finite temperature. Note how increasing the temperature of the star inflates its radius for a given mass as a result of the effect of thermal pressure. Figure~\ref{fig:mr} also illustrates the properties of three classical white dwarfs, whose masses and radii were independently determined due to their belonging to binary or triple star systems (in general the mass and radius of a white dwarf cannot be independently measured). These measurements match the theoretical mass--radius relations. The same conclusion has been reached for many other stars using other observational techniques that can yield independent mass and radius measurements for specific stellar systems, such as astrometric microlensing and the characterization of eclipsing binaries. In fact, the mass--radius relationship is now well established to the extent that determining one, the mass or the radius, is generally considered to directly provide an accurate determination of the other, and vice-versa. Similarly, a measurement of the surface gravity $g = GM / R^2$, a convenient quantity in the analysis of stellar spectra, can be used to uniquely determine the mass and radius of a white dwarf.

\section{White dwarfs as the fossil records of stellar evolution}\label{sec:evol}
Having explored the basic properties and nature of white dwarfs, it is now time we turn to their origins. White dwarfs represent the end point in the evolution of stars with initial masses lower than approximately $10\,M_{\odot}$. This encompasses 98\% of the stars in our Milky Way, with the remainder collapsing into neutron stars or black holes. Today, around 6\% of stars in our Galaxy are white dwarfs. The path to this final state involves a sequence of nuclear fusion processes that are summarized below. We first focus on the formation of white dwarf cores, where most of the mass is contained, in Sections~\ref{sec:core1} and~\ref{sec:uncertain_composition} before discussing the outer layers in Section~\ref{sec:envelope}.

\subsection{The formation of white dwarf cores} 
\label{sec:core1}
The first step in the creation of a white dwarf core is when hydrogen undergoes fusion into helium, a process that begins in the main sequence stage and continues into the subgiant and red giant branch (RGB) phases. Stars with a mass inferior to about $0.5\,M_{\odot}$ do not continue further down the periodic table to ignite helium fusion. These stars generally remain on the main sequence for durations exceeding the age of the universe and therefore cannot a priori contribute to the current white dwarf population. Nonetheless, helium-core white dwarfs, stars that did not undergo further nuclear burning, do exist. The formation of these rare objects is generally attributed to evolution within binary systems that experience mass-transfer episodes. Indeed, the majority of these helium-core white dwarfs are found in binary systems. These stars have masses below $\simeq 0.45\,M_{\odot}$, and the most extreme specimens, with masses inferior to $\simeq 0.25\,M_{\odot}$, are known as extremely low-mass white dwarfs (ELMs).

Excluding the progenitors of helium-core white dwarfs, the second step in the creation of a white dwarf core starts when a star progresses to ignite helium burning. During the horizontal branch phase, helium is burned in a convective core surrounded by a stable helium envelope. The helium-fusing core produces carbon-12 through the triple-$\alpha$ reaction and oxygen-16 via the $\alpha$-capture reaction $^{12}{\rm C}(\alpha,\gamma)^{16}{\rm O}$, ultimately leading to the formation of a carbon--oxygen core. This region is homogeneous (i.e., the carbon-to-oxygen ratio is the same everywhere) because it was produced in a convective environment where fluid motions efficiently mix the different chemical species. After the horizontal branch, the star will continue burning helium during the asymptotic giant branch (AGB) phase. AGB stars are characterized by a sandwich-like structure, with a carbon--oxygen core (the remnant of the core helium burning phase) surrounded by helium-burning shell, a hydrogen burning shell, and an inert intershell in between. Evolved AGB stars undergo periodic instabilities known as ``thermal pulses''. This phenomenon is due to the mismatch between the helium production rate by the hydrogen-burning shell and the rate at which helium is fused in the helium-burning shell underneath. Helium thus piles up and is compressed to become moderately degenerate. This catastrophically increases the helium burning rate as the helium-burning shell can now heat up without significantly expanding (remember that thermal pressure is negligible in a degenerate gas). This is also known as a helium shell flash.

\begin{figure}
\centering
\includegraphics[width=0.6\textwidth]{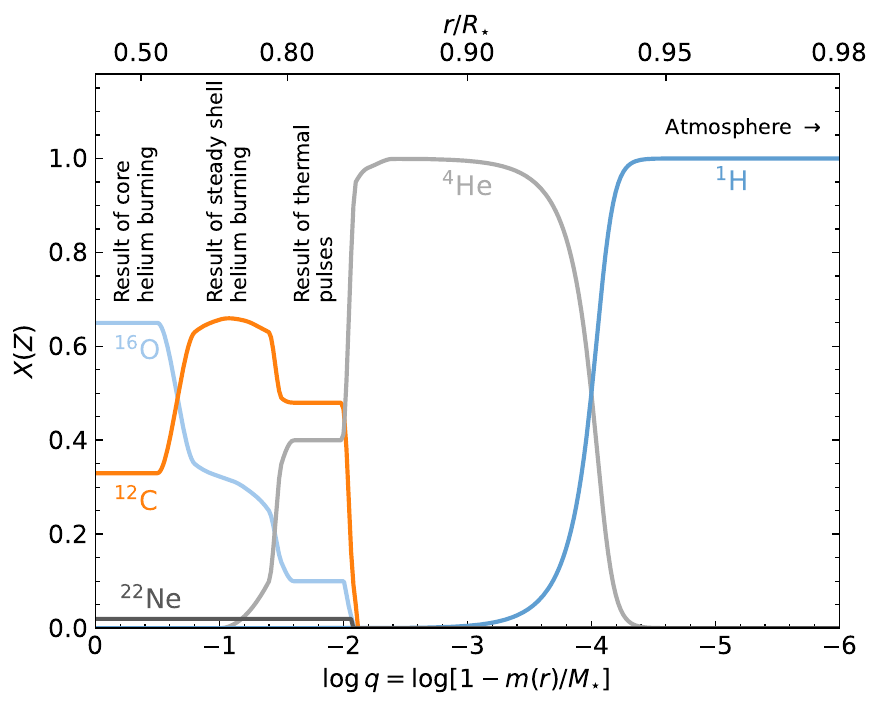}
\caption{Representative chemical composition profile of a typical white dwarf. The abundances of each species are given as mass fractions. The center of the star is at the left, and the surface is towards the right. $m(r)$ is the mass enclosed within a sphere of radius $r$; hence, $q$ corresponds to the mass fraction outside of radius $r$. For example, 90\% of the white dwarf's mass is located to the left of $\log q = -1$. It is useful to think of a white dwarf as being divided into three parts: a core (in most cases made of carbon and oxygen), atop of which sits a helium--hydrogen envelope, and an atmosphere (the very thin layer at the surface that is observable, $\log q \lesssim -14$). Note how the envelope makes up only 1\% of the mass but accounts for 15\% of the radius given its relatively low density.}
\label{fig:comp}
\end{figure}

The helium-burning phases detailed in the previous paragraph ultimately result in the formation of a carbon--oxygen core that comprises approximately 99\% of the mass of the soon-to-be-born white dwarf. As indicated in Figure~\ref{fig:comp}, this core bears the imprints of the different helium-burning phases. A large portion of the mass is contained in a central homogeneous region formed during the core helium burning phase, and above that we find a more complex composition profile resulting from the AGB evolution. As we will see in Section~\ref{sec:uncertain_composition}, the core composition profile remains uncertain, reflecting unresolved questions related to the helium-fusing phases of stellar evolution.

The carbon--oxygen core also harbor a small trace of neon-22, representing 1--2\% of the white dwarf's mass.\footnote{This fraction corresponds in very good approximation to the initial metallicity of the progenitor star.} This isotope is formed during the helium-burning phases, through the capture of alpha particles by nitrogen-14, which itself is a byproduct of the hydrogen-burning CNO cycle. Although neon-22 is only a trace species in carbon-oxygen white dwarfs, its presence is noteworthy due to its neutron-rich composition, featuring 12 neutrons and 10 protons, unlike carbon-12 and oxygen-16 which have equal numbers of neutrons and protons. This excess of neutrons means that more mass is necessary to achieve the same pressure, which as we have seen is primarily sustained by electron degeneracy pressure. Neon-22 requires 2.2 nucleons per electron, compared to 2.0 nucleons per electron for carbon-12 and oxygen-16. This has interesting implications for the subsequent evolution of the white dwarf (Section~\ref{sec:chemevol}), which are the subject of active research and have yet to be fully understood.

Finally, stars with an initial mass close to the $10\,M_{\odot}$ upper limit for white dwarf progenitors can go one step further in their nuclear burning. These objects can reach the super asymptotic giant branch (SAGB), where the temperature is high enough to ignite carbon and ultimately produce an oxygen--neon core. This core is dominated by oxygen-16 and neon-20, but also contains significant traces of carbon-12, neon-22, sodium-23, and magnesium-24. Oxygen--neon white dwarfs are rare. A white dwarf needs to have a mass $\gtrsim 1.05\,M_{\odot}$ (these stars are known as ultramassive white dwarfs) to have evolved through the SAGB phase, and only $\sim 2\%$ of white dwarfs reach that mass. In fact, the fraction of white dwarfs hosting oxygen--neon cores might be even smaller, as there is now convincing evidence that some ultramassive white dwarfs actually harbor carbon--oxygen core, contrary to what standard (single-star) stellar evolution theory predicts \citep{cheng2019}.

\subsection{The uncertain core composition profile}
\label{sec:uncertain_composition}
The white dwarf core composition profile shown in Figure~\ref{fig:comp} left of $\log q=-2$  aligns with the predictions of current stellar evolution theory, but these predictions remain uncertain. First, there is the question of the $^{12}{\rm C}(\alpha,\gamma)^{16}{\rm O}$ reaction rate. As we have seen, this reaction is responsible for the production of oxygen-16 during the helium-burning phases, and as such, it largely controls the carbon-to-oxygen ratio of white dwarf cores. However, the cross section of $^{12}{\rm C}(\alpha,\gamma)^{16}{\rm O}$ at the temperatures relevant for helium burning in stars is very small, making its empirical determination challenging. Decades of experimental work have significantly narrowed down the uncertainties on the $^{12}{\rm C}(\alpha,\gamma)^{16}{\rm O}$ reaction rate, but these reduced uncertainties still imply a relatively wide range of possible oxygen abundances. Varying the reaction rate within the $2\sigma$ experimental confidence interval results in changes of the order of $\pm 0.1$ in the oxygen mass fraction at the center of white dwarfs \citep{chidester2022}. This $\simeq 15\%$ uncertainty has important consequences for the subsequent evolution of the white dwarf.

Second, we still do not know how large the central oxygen-rich region really is. The mass extent of this homogeneous region corresponds to the mass extent of the former helium-burning convective core, where species were efficiently mixed by convective eddies. Standard stellar evolution provides a simple prescription, the Schwarzschild criterion, to define the location of a convective boundary. According to this criterion, the star is convective in regions that satisfy the condition $\nabla_{\rm rad} > \nabla_{\rm ad}$, where $\nabla_{\rm rad}$ is the so-called radiative gradient,
\begin{equation}
    \nabla_{\rm rad} = \frac{3}{16 \pi a c G} \frac{\kappa L P}{m T^4},
    \label{eq:nablarad}
\end{equation}
with $a$ the radiation constant, $\kappa$ the opacity of the gas, and $m$ the mass enclosed within the radius at which $\nabla_{\rm rad}$ is calculated. $\nabla_{\rm ad}$ is the adiabatic gradient, the temperature variation in a fluid element undergoing a change in pressure at constant entropy,
\begin{equation}
    \nabla_{\rm ad} = \left( \frac{\partial \ln T}{\partial \ln P} \right)_s.
\end{equation}
The value of $\nabla_{\rm ad}$ depends on the equation of state of the fluid (e.g., $\nabla_{\rm ad} = 0.4$ for an ideal gas without molecules). Despite the pervasiveness of the Schwarzschild criterion in stellar evolution models, it is well established that mixing due to convection can occur passed the Schwarzschild boundary, in regions where $\nabla_{\rm rad} < \nabla_{\rm ad}$. In fact, establishing just how far this convective boundary mixing extends is an area of active study for all sorts of stellar environments, from the hydrogen-burning cores of main-sequence cores to the convective envelopes of white dwarfs. This problem turns out to be particularly challenging in the case of convective helium-burning cores. A complication arises due to the production of carbon and oxygen in the convective core, which are much more opaque than the helium located above. This tends to create a discontinuity in the $\kappa$ profile, which makes the convective boundary unstable. Indeed, any amount of extra mixing beyond the convective boundary would push high-opacity material in the helium-rich layer above, thereby increasing its $\nabla_{\rm rad}$ and making it convective. But to due the particular interactions between the different thermodynamic quantities at play, this outward extension eventually splits the convective region in two. This outcome is generally considered to be unphysical and many numerical schemes have been devised to circumvent this problem. These different prescriptions ultimately lead to different white dwarf composition profiles, and it remains unclear which of the multiple approaches implemented in various stellar evolution codes should be preferred. Note that uncertainties on core boundary mixing do not only affect the size of the oxygen-rich region, but also its carbon-to-oxygen ratio.

Third, the region formed during the thermally-pulsing AGB phase ($-2 \lesssim \log q \lesssim -1.5$ in Figure~\ref{fig:comp}) is also fraught with uncertainty. Successive thermal pulses convert helium into carbon and oxygen, and homogenize the shell due to the development of a convection zone. This is thought to ultimately result in the formation of a complex double-layered structure at the bottom of the helium envelope. However, the size and composition of this region are very uncertain. They are affected by the number of thermal pulses experienced by the star, which itself depend on the rate at which the star loses mass during the thermally-pulsing AGB phase and on the uncertain extent of convective boundary mixing in the flash-driven convection zone.

These different sources of uncertainties are inconvenient in that they currently limit our ability to precisely model white dwarfs. Yet, the fact that white dwarfs are sensitive to these earlier processes also represents a unique opportunity. In particular, asteroseismology, the study of pulsations in stars, is emerging as a promising tool to semi-empirically infer the composition profiles of the interiors of white dwarfs \citep{giammichele2018}. The hope is that white dwarfs can then play the role of stellar fossils that, when correctly examined, shed light on earlier evolutionary phases to constrain the $^{12}{\rm C}(\alpha,\gamma)^{16}{\rm O}$ reaction rate \citep{chidester2022} or the physics of convective boundary mixing.

\subsection{The envelope}
\label{sec:envelope}
We have discussed at length the formation of the carbon--oxygen core, but what about the helium and hydrogen layers above (Figure~\ref{fig:comp})? These layers represent only $\sim 1\%$ of the total mass, but they play a fundamental role in our understanding of white dwarfs. Not only is the only observable region of the star (the atmosphere, see Section~\ref{sec:spectro}) located there, but also the hydrogen/helium envelope largely controls the subsequent evolution of the star by mediating the flow of energy between the core and the exterior.

Standard stellar evolution predicts an helium envelope making up $\sim 1\%$ of the star's total mass ($M_{\rm He} \sim 10^{-2} M_\star$) and a much thinner superficial hydrogen layer, $M_{\rm H} \sim 10^{-4} M_\star$, where the exact numbers depend on the metallicity and stellar mass (thinner hydrogen and helium layers with increasing mass). Asteroseismological studies largely confirm that these predictions are correct for typical white dwarfs \citep{giammichele2022}. The size of the hydrogen layer is determined by the maximum extent of the hydrogen-burning region. If an hydrogen envelope is too thick, the conditions for nuclear burning are met at its base, and the hydrogen layer then shrinks at the expense of the helium envelope. This residual burning is negligible for most white dwarfs, but not for low-metallicity white dwarfs \citep{bertolami2013}. That being said, we also know from observations (see Section~\ref{sec:chemevol}) that $\sim 25\%$ of white dwarfs must have a hydrogen layer that is much thinner than this canonical value of $M_{\rm H} \sim 10^{-4} M_\star$. The main evolutionary pathway associated with the formation of these so-called hydrogen-deficient white dwarfs is the born-again scenario \citep{werner2006}. According to this scenario, a post-AGB star or a white dwarf experiences a late helium-shell flash after having left the AGB. The envelope then rapidly expands and becomes convective. This convection zone engulfs the residual hydrogen and transports it to deeper (and hotter) regions where it can be burnt. When this final burning episode is completed, the star contracts into a white dwarf, now with almost no hydrogen left (as little as $M_{\rm H} \sim 10^{-12} M_\star$ in some cases).

\section{The evolution of white dwarfs}
\label{sec:cooling}
Having discussed in some detail the evolutionary history leading to the formation of a white dwarf, we now turn our attention to their subsequent evolution upon reaching the white dwarf stage. This has been a subject of active research for decades. Given the extensive nature of this topic, it is covered in greater depth elsewhere in this Encyclopedia. However, a concise overview is essential here to provide context and framework for the discussions that follow. This section aims to offer a brief summary of the main themes and processes involved in white dwarf cooling. We will also take a closer look at the physical conditions that characterize white dwarfs in Section~\ref{sec:rhoT}.

\subsection{White dwarfs as cooling embers}
\label{sec:cooling51}
\begin{figure}
\centering
\includegraphics[width=0.6\textwidth]{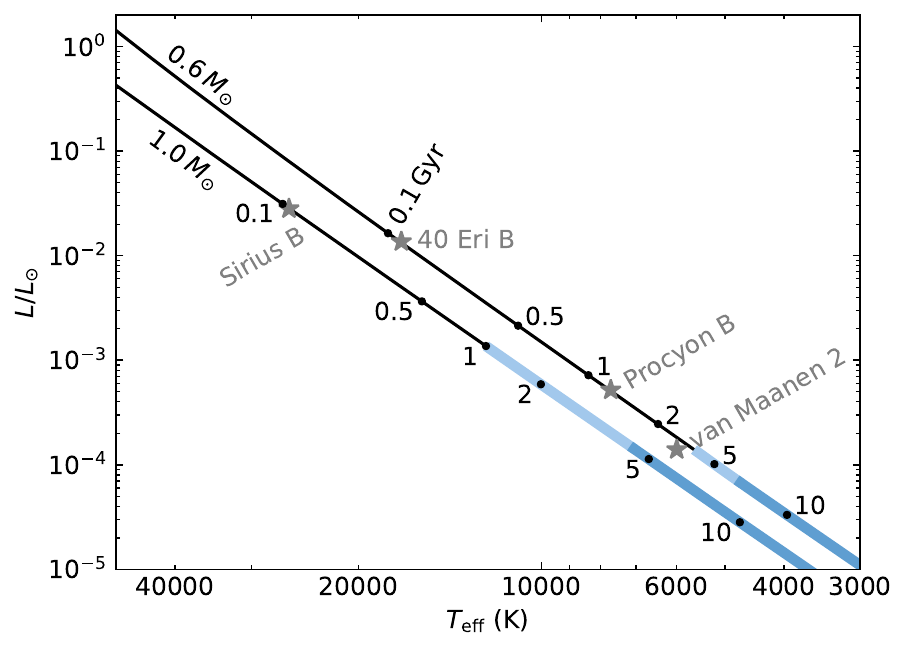}
\caption{Theoretical white dwarf cooling tracks for carbon--oxygen white dwarfs with thick hydrogen envelopes and masses of 0.6 and 1.0$\,M_{\odot}$ (top and bottom, respectively). Small black circles with adjacent labels indicate cooling times in Gyr. The portions of the cooling tracks colored in light blue correspond to white dwarfs in the process of crystallizing, while the dark blue portions mark complete core crystallization. Some well-known white dwarfs are shown in grey.}
\label{fig:cooling}
\end{figure}

The vast majority of white dwarfs do not experience any nuclear burning.\footnote{Notable exceptions include accreting white dwarfs in binary systems and moderate residual hydrogen burning in white dwarfs descending from low-metallicity progenitors.} Deprived of this energy source, white dwarfs behave as stellar embers that can only cool down. This makes white dwarf evolution relatively simple compared to other types of stars. As shown in Figure~\ref{fig:cooling}, white dwarfs simply follow a straight trajectory in a luminosity--effective temperature diagram (Hertzsprung--Russell diagram), becoming ever cooler and fainter. This evolutionary pathway is often called a cooling track.

The fact that the cooling tracks in Figure~\ref{fig:cooling} are nearly perfectly linear in this log-log diagram implies that the stars' radii remain virtually constant as they cool down (see Equation~\ref{eq:sb}). This is of course a direct consequence of the degenerate nature of white dwarfs, which largely prevents further gravitational contraction (see also the small difference in radius between white dwarfs with identical masses but different temperatures in Figure~\ref{fig:mr}). This is an important property, as it implies that white dwarfs cannot tap into gravitational contraction to maintain their luminosity. Very young and hot white dwarfs do experience a non-negligible decrease ($\sim$\,50\%) of their radii in their first few Myr of evolution. However, this contraction mostly takes place in their non-degenerate outer envelopes. As we have seen (Figure~\ref{fig:comp}), the envelope represents a very small fraction of the white dwarf's total mass, meaning that the gravitational energy released by this process is only a minor source of energy.

Most of a white dwarf's thermal energy is stored by the ions in the core, where $\simeq 99\%$ of the mass is. The electrons contribute little to the total thermal reservoir, since as degenerate particles, they already lie in the lowest energy states available to them. White dwarf cooling is then mostly a matter of emptying that ionic thermal energy reservoir outside the star. For most of the white dwarf's evolution, this is ultimately done by photons escaping from the surface, but in the early stages of the cooling process, neutrinos produced and escaping from the core are the main energy sink. 

There are two main questions when it comes to white dwarf cooling: how large is the thermal reservoir that is slowly leaking, and how fast is this leaking taking place. The first question is closely connected to the discussion of Section~\ref{sec:uncertain_composition}. If we assume for simplicity that the core is made of non interacting ions, classical thermodynamics tells us that the heat capacity of each ion in the core is given by $\frac{3}{2} k_B$. Since a core of a given mass will contain a different number of ions depending on its composition, it follows that the heat capacity and stored thermal energy is dependent on the core composition. For example, a more oxygen-rich core will have fewer ions than a more carbon-rich core, and all else being equal, cool down more quickly as a result. As we have seen, uncertainties in pre-white dwarf evolution imply that the core composition remains only loosely constrained. 

The second question is a complicated one that depends on a wide range of interesting physical processes, including thermal conduction in dense plasmas, convection, radiative opacities in the atmosphere, and the physics of phase transitions. This last point is particularly noteworthy and merits further discussion, even in this brief overview of white dwarf cooling. As the ions in the core gradually lose their thermal energy, they become more and more influenced by the Coulomb interactions with their neighbours. The ratio of their electrostatic interaction energy to their kinetic energy gradually increases. This is measured by the Coulomb coupling parameter,
\begin{equation}
    \Gamma = \frac{Z^2 e^2}{a_i k_B T},
    \label{eq:gamma}
\end{equation}
where $Z$ is the ionic charge and $a_i$ is the average interionic distance. The plasma freezes into a solid state when $\Gamma$ reaches a critical threshold ($\Gamma=175$ for the simple case of a plasma with just one ionic species). This liquid--solid phase transition is known as core crystallization.\footnote{Crystallized white dwarf cores are sometimes referred to as ``diamonds''. While both diamonds and frozen white dwarf cores contain crystallized carbon, there are of course important differences between the two materials. Both crystals have a different lattice structures; a diamond is made of carbon atoms interacting through covalent bonds, while repulsive Coulomb forces are responsible for white dwarf crystallization; and white dwarf cores always contain large amounts of oxygen alongside carbon.} Core crystallization unfolds over Gyr timescales (see the light blue highlighting in Figure~\ref{fig:cooling}) as a crystallization front slowly moves outward starting from the center of the star. This inside-out progression is the result of a weak temperature gradient (Section~\ref{sec:rhoT}) and strong density gradient in the core. The center being more dense, the Coulomb interactions are stronger there ($a_i$ is smaller in Equation~\ref{eq:gamma}), which favors crystallization. Importantly for white dwarf evolution, this transition is accompanied by the release of latent heat, which, as a new energy source, slows down the cooling process. Ultimately, this transformation into a solid state accelerates the cooling of the coolest white dwarfs by dramatically decreasing the heat capacity of the core. This is known as Debye cooling.

\subsection{Chemical evolution}
\label{sec:chemevol}
As a consequence of being compact objects, white dwarfs have an intense surface gravity of $\sim 10^8\,{\rm cm\,s}^{-2}$ ($\log g = 8$, compared to $\log g=4.4$ for the Sun). This implies that gravitational settling is particularly efficient, with heavier elements sinking down and lighter ones floating up. However, this simple picture must be incomplete, since as we will see in Section~\ref{sec:spectro}, elements heavier than hydrogen are often detected at the surface of white dwarfs. Additional physical mechanisms compete with gravitational settling so that the very thin observable layer at the surface can take different compositions. Interestingly, these transport mechanisms change in importance during the evolution of white dwarfs, such that the surface composition of a given white dwarf changes over the course of its evolution. This is often referred to as spectral evolution.

The general picture that has emerged from decades of empirical and theoretical studies is as follows. As we have seen in Section~\ref{sec:envelope}, standard stellar evolution theory predicts that white dwarfs have a hydrogen content corresponding to $M_{\rm H} \sim 10^{-4} M_{\star}$. This is generally referred to as a ``thick'' hydrogen layer, and with that amount of hydrogen no internal transport mechanism can alter the composition of the surface layer. Hence, these white dwarfs, which represent approximately 75\% of the white dwarf population maintain a pure-hydrogen atmosphere throughout their evolution (assuming that no external accretion contaminates their surface). The other 25\% are more interesting. They enter the white dwarf cooling track with much less hydrogen, probably following a late helium shell flash (Section~\ref{sec:envelope}). Initially, at very high temperatures (down to $\simeq 75{,}000\,$K), their atmospheres are rich in carbon and oxygen. These heavy elements can remain at the surface despite the incessant pull of gravitational settling because of competition from stellar winds. Eventually, the winds fade and gravitational settling can operate more freely. The small amount of residual hydrogen contained in these stars then rises to the surface, and most display a hydrogen-dominated atmosphere by the time they reach $\simeq 30{,}000\,$K. The amount of hydrogen required to complete this transformation into a hydrogen-atmosphere white dwarfs is surprisingly small, with $M_{\rm H} \sim 10^{-12} M_{\star}$ thought to be sufficient. As the cooling process continues, these stars eventually recover a hydrogen-deficient atmosphere. For stars with very little hydrogen, this is done through the process of convective dilution in the $30{,}000\,{\rm K} \gtrsim T_{\rm eff} \gtrsim 15{,}000\,{\rm K}$ range. The thin superficial hydrogen layer is eroded from beneath by convective motions in the much thicker helium layer. For stars with more hydrogen, convection instead develops in the hydrogen layer, thereby mixing it with the thicker helium mantle underneath provided that the hydrogen layer is not too thick ($M_{\rm H} \lesssim 10^{-6} M_{\star}$). This takes place below $15{,}000\,$K, with more massive hydrogen layers necessitating cooler temperatures (the convection zone deepens as the white dwarf cools). Spectral evolution contains many more complexities and unsolved riddles than what has been presented in this brief outline: the interested reader should consult \cite{bedard2024} for a more complete up-to-date overview of this subfield.

The surface layers are not the only ones to undergo compositional changes. First, gravitational settling also operates in the core. For example, in a carbon--oxygen white dwarf neon-22 slowly diffuses downward as a result of its higher mass-to-charge ratio, a process often referred to as neon-22 sedimentation. This is noteworthy because the transport of heavier species toward the center releases gravitational energy that can materially slow down the cooling process by providing a new energy source to the star. By the time neon-22 diffusion stops because of the core being completely crystallized, this process can induce a cooling delay of a few hundreds of Myr (meaning that the white dwarf would have reached a given temperature/luminosity hundreds of Myr earlier if it were not for neon-22 sedimentation). A more spectacular chemical transformation takes place during the crystallization process itself. The newly formed solid phase generally does not have the same composition as the coexisting liquid. This fractionation process is similar to the formation of sea ice, where salt is largely expelled from the solid phase. In white dwarfs, this chemical separation rearranges the core composition profile in a way that can release copious amounts of gravitational energy. This is enough to delay the cooling process by many hundreds of Myr for the vast majority of white dwarfs, and for some specific core compositions this can extend to several Gyr \citep{blouin2021}.

\subsection{A detailed look at the stratification}
\label{sec:rhoT}
\begin{figure}
\centering
\includegraphics[width=0.6\textwidth]{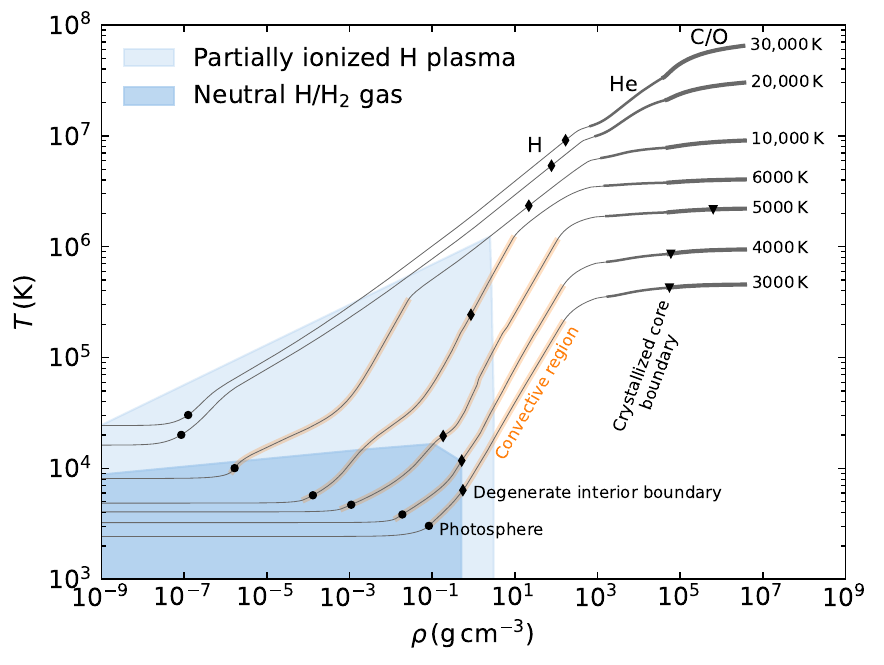}
\caption{White dwarf density--temperature profiles. These profiles were calculated for a $0.6\,M_{\odot}$ white dwarf with a thick hydrogen layer ($M_{\rm H}/M_{\star}=10^{-4}$) for effective temperatures ranging from 30{,}000\,K (top profile) to 3000\,K (bottom profile). The center of the star is on the right; the surface is on the left. The thickness of the lines indicates the dominant atomic constituent (hydrogen, helium, or carbon/oxygen). The white region corresponds to a fully ionized plasma, while the different shadings indicate a partially ionized plasma (light blue) or a neutral hydrogen gas (darker blue). The regions highlighted in orange are convective. A small black circle on each profile marks the location of the photosphere, a diamond indicates the outer boundary of the degenerate interior, and a triangle delimits the extent of the solid core for objects that are cool enough to have undergone crystallization.}
\label{fig:Trho}
\end{figure}

In this section, we focus on the specific physical conditions within white dwarfs, examining their internal structure from the dense core all the way to the tenuous atmosphere at their surface. Figure~\ref{fig:Trho} shows density--temperature profiles for a standard $0.6\,M_{\odot}$ carbon--oxygen core white dwarf with a thick hydrogen layer ($M_{\rm H}/M_{\star}=10^{-4}$) at different effective temperatures. The thickness of the lines indicate the dominant atomic constituent, with carbon and oxygen in the dense core on the right (thickest lines), followed by helium, and finally hydrogen on the left towards the surface (thinnest lines). Note that this representation greatly expands the outer layers of the star: in Figure~\ref{fig:comp} we saw that the carbon--oxygen core represents 85\% of the radial extent and 99\% of the mass of the star. 

The densities and temperatures encountered in white dwarfs span many orders of magnitude, and as such, matter is found in different states. At high temperatures and/or high densities, we have a completely ionized plasma. This corresponds to the region with no color shading in Figure~\ref{fig:Trho}. If the plasma is both very dense and not too hot, we have seen in Section~\ref{sec:cooling51} that the plasma freezes into a solid state. This liquid--solid boundary is indicated by triangular symbols in Figure~\ref{fig:Trho}. At the other extreme, at very low temperatures and densities close to the surface, we find a neutral hydrogen gas (region shaded in dark blue). For the coolest white dwarfs, this neutral hydrogen gas is dominated by molecular hydrogen. Note that Figure~\ref{fig:Trho} only considers the case of a thick hydrogen layer: white dwarfs with much less hydrogen can instead have a neutral helium gas close to the surface. Finally, there is a region of partial ionization (region shaded in light blue) in between the completely ionized plasma in the interior and the neutral hydrogen gas close to the surface. This region of partial ionization is associated with a convection zone. This is due to the strong opacity increase that makes convection the most efficient way of transporting to the surface heat from the thermal reservoir in the core. This particular region of the star is also responsible for the driving of pulsations in pulsating white dwarfs, which is discussed in more details elsewhere in this Encyclopedia. Also shown in Figure~\ref{fig:Trho} is the boundary of the degenerate interior (diamond symbols), defined here as the point where the temperature is inferior to the local Fermi temperature, the temperature at which thermal effects are comparable to quantum effects due to the Pauli exclusion principle. We can see that both the helium layer and the carbon--oxygen core are in this electron degenerate state.

We mentioned convection as an energy transport mechanism in the previous paragraph, but this is just one out of four ways of transporting energy out of the core in white dwarfs. First, we already saw earlier how neutrino cooling be an important energy sink for very hot white dwarfs. Second, radiative transport is the dominant energy transport mechanism for the outer layers of the star above the convection zone (if there is one). Radiative transport is negligible in the interior because radiative opacities are too high for this to be an efficient transport mechanism. Third, electron thermal conduction dominates energy transport for the bulk of the star. In fact, it is so efficient in dense degenerate plasmas that the cores of white dwarfs are nearly isothermal (notice how shallow the slope of the density--temperature profiles becomes at very high densities in Figure~\ref{fig:Trho}). The region where the electron thermal conductivity drops (in the vicinity of the degenerate interior boundary) acts as a sort of bottleneck for white dwarf cooling.

\section{White dwarfs demographics}\label{sec:demographics}
We are today far from the early years of white dwarf science where only a handful of objects were known. Over 350{,}000 white dwarf candidates have been identified with high confidence based on their distances and magnitudes \citep{gentile2021}, and tens of thousands of stars are confirmed to be white dwarfs given the appearance of their optical spectra \citep{dufour2017}. These large numbers enable statistical studies of the properties of white dwarfs. However, interpreting these statistics can be challenging due to biases in sample selection (e.g., brighter stars are easier to detect than fainter ones), a common issue in astronomical studies. To mitigate this, we can focus on a volume-complete sample, where all white dwarfs within a given radius from the Sun are analyzed. Such a sample, with a very high degree of completeness, is now available up to 40\,pc. We take a closer look at this sample below.

\subsection{The mass and temperature distributions}

\begin{figure}
\centering
\includegraphics[width=0.6\textwidth]{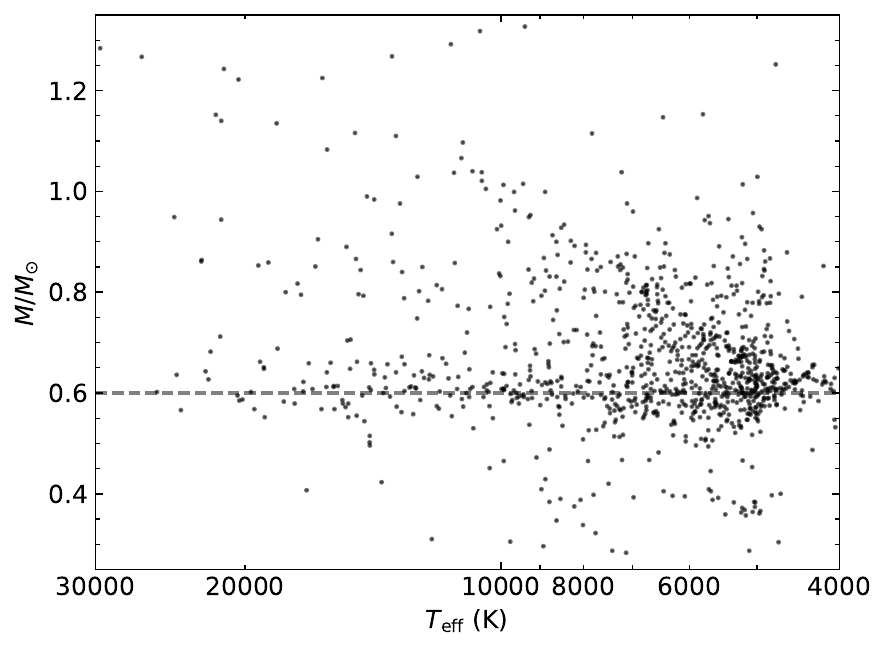}
\caption{Distribution of white dwarfs in the mass--effective temperature plane. Every dot represents a star from the 40\,pc volume-complete sample of \cite{obrien2024}. A dashed gray line marks $0.6\,M_{\odot}$ to guide the eye.}
\label{fig:mteff}
\end{figure}

\begin{figure}
\centering
\includegraphics[width=0.6\textwidth]{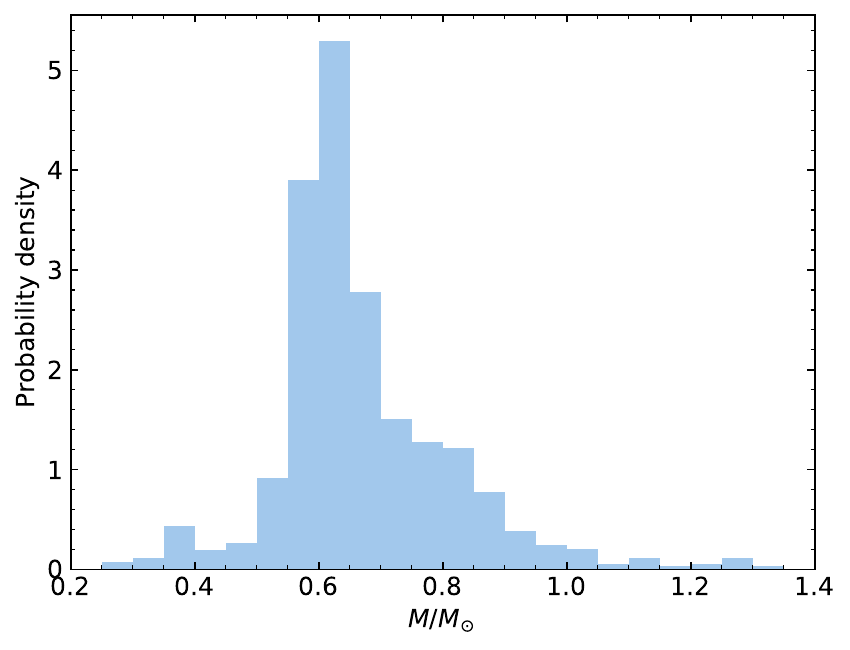}
\caption{The mass distribution of white dwarfs. The histogram is normalized so that it integrates to one. The same sample as in Figure~\ref{fig:mteff} was used to generate this figure.}
\label{fig:mdis}
\end{figure}

Figure~\ref{fig:mteff} shows the distribution in the mass--effective temperature plane of all known white dwarfs within 40\,pc of the Sun.\footnote{It is beyond the scope of this chapter to explain how these masses and temperatures are measured (this is discussed at length in the chapter on white dwarf observations in this Encyclopedia). For the purpose of this discussion, it suffices to know that these quantities can be measured with a precision of a few percent for most white dwarfs.} On a diagram like this one, white dwarfs evolve from left to right (decreasing temperature) at constant mass. The most striking aspect of Figure~\ref{fig:mteff} is the clustering of stars around $0.6\,M_{\odot}$, indicating that most white dwarfs have masses close to $0.6\,M_{\odot}$. This fact is more explicitly illustrated in Figure~\ref{fig:mdis}, where the white dwarf mass distribution is shown. The distribution shows a sharp peak centered at $0.61\,M_{\odot}$. There is also a broad shoulder at larger masses (whose exact origin remains an open question), but overall the mass distribution is surprisingly narrow given the wide range of stellar masses spanned by stars that eventually become white dwarfs. This question is explored in greater details elsewhere in this Encyclopedia (see the chapter on the initial--final mass relation), but in short, more massive stars lose more mass than less massive ones during the late stages of stellar evolution. This implies that white dwarf progenitors with very different initial masses converge to comparable masses by the time they reach the white dwarf state. For example, a $1\,M_{\odot}$ star will evolve to become a $\simeq 0.57\,M_{\odot}$ white dwarf, while a star twice that mass ($2\,M_{\odot}$) yields a barely more massive white dwarf ($\simeq 0.65\,M_{\odot}$).

Another remarkable aspect of Figure~\ref{fig:mteff} concerns the distribution of white dwarfs along the effective temperature axis. In particular, notice how the number of white dwarf per 1000\,K interval increases with decreasing temperature (at least down to $\simeq 5000\,$K). In an unbiased white dwarf sample, hot white dwarfs are much more rare than cool white dwarfs. The fundamental reason for this distribution is shown in Figure~\ref{fig:cooling}: the cooling rate of a white dwarf generally decreases with time. A $0.6\,M_{\odot}$ white dwarf needs more time to cool down from $T_{\rm eff} = 100{,}000$\,K all the way to 6000\,K, than for cooling down from 6000\,K to 5000\,K. Part of this behavior is due to specific physical processes that slow down the cooling process (e.g., crystallization), but a simple, general explanation is that the cooler the surface becomes, the more slowly the star can radiate away its energy (Equation~\ref{eq:sb}).

\subsection{The white dwarf luminosity function}
Interestingly, this trend of increasing numbers of white dwarfs with decreasing temperature eventually comes to an end. Very few white dwarfs cooler than 4000\,K are known, even beyond the 40\,pc sample. Of course, such cool white dwarfs are harder to identify given their intrinsic faintness, but this fact alone cannot explain the clear decrease in the number of white dwarfs below $\simeq 5000\,$K in the almost complete 40\,pc sample of Figure~\ref{fig:mteff}. The solution to this conundrum is simply that the disk of our Galaxy is still too young to have produced older white dwarfs. A $0.6\,M_{\odot}$ white dwarf had to go through $\simeq 4\,$Gyr of stellar evolution before even reaching the white dwarf phase, and it then needs another $\simeq 6\,$Gyr to cool down to 5000\,K. Given that the galactic disk is estimated to be $\simeq 10\,$Gyr old, this naturally explains the paucity of cooler stars in Figure~\ref{fig:mteff}. 

\begin{figure}
\centering
\includegraphics[width=0.6\textwidth]{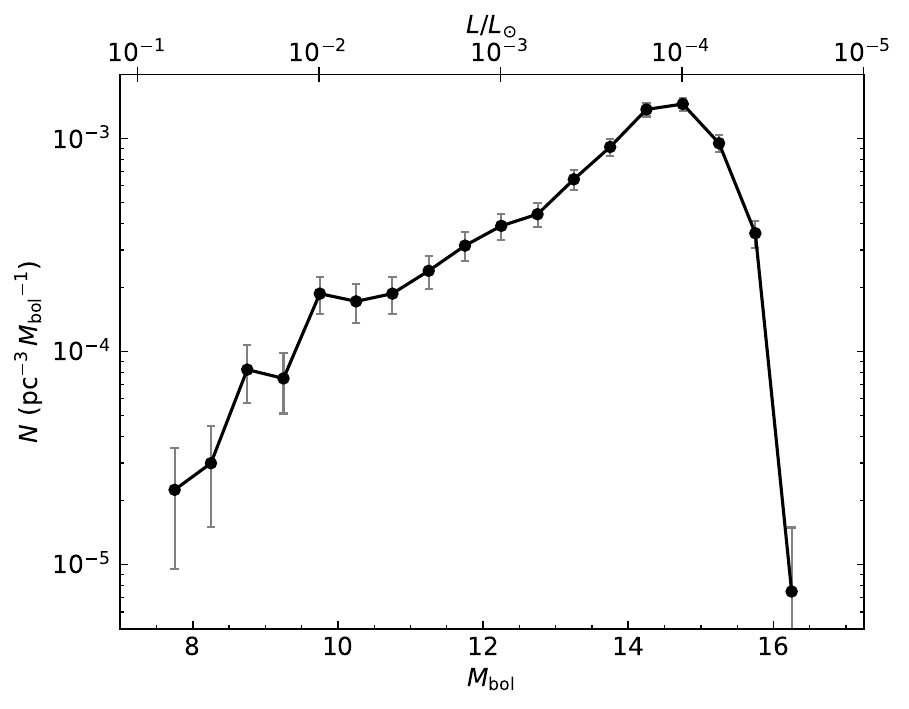}
\caption{The white dwarf luminosity function. The curve shows the number of white dwarfs per luminosity bin (or, similarly, per bolometric magnitude bin, $M_{\rm bol} = 4.74 - 2.5 \log_{10} (L/L_{\odot})$) and normalized by volume. The gray error bars show the $1\sigma$ uncertainties based on Poisson statistics in each bin. The same sample as in Figure~\ref{fig:mteff} was used to generate this figure.}
\label{fig:lumf}
\end{figure}

This property of the temperature distribution is not merely a curiosity. It implies that white dwarfs can be used as cosmic clocks to constrain the age of the galactic disk and other stellar populations. The oldest white dwarf of any given stellar population yields a firm lower limit on the age of that population. The analysis of white dwarfs with the aim of constraining ages is known as white dwarf cosmochronology, and it represents one of the key motivations underpinning current-day studies of white dwarfs. A common way to characterize this downturn in the white dwarf population at old ages is through the luminosity function (Figure~\ref{fig:lumf}). It consists of a volume-normalized luminosity (or bolometric magnitude) distribution of white dwarfs. In other words, it shows for a cubic parsec of space how many white dwarfs there are within a given luminosity bin. Figure~\ref{fig:lumf} shows an increasing number of white dwarfs with decreasing luminosity, which reflects the decreasing cooling rate. This is followed by an abrupt cut-off around $\log L/L_{\odot}=-4.5$ that marks the finite age of the galactic disk \citep{winget1987}. Beyond age-dating applications, the luminosity function is also a useful tool to test and constrain white dwarf cooling models as its detailed shape is affected by the various physical ingredients (e.g., crystallization physics, opacities) that control the cooling rate.  

\section{White dwarf atmospheres}
\label{sec:spectro}
Being only $\sim 100\,$m thick, the atmosphere of a white dwarf represents a tiny fraction of the star's total radius and mass. Nonetheless, the importance of this minuscule layer is hard to overstate, as it is the only observable portion of the star. 

The light emerging from a white dwarf mostly comes from a region known as the photosphere (marked by circles in Figure~\ref{fig:Trho}), which is defined as the layer where 50\% of the photons escape the atmosphere without further absorption or scattering. While the photosphere represents the \textit{typical} depth from which photons that we observe come from, it is important to note that when a white dwarf is observed, we are effectively integrating along a line of sight and many other layers contribute to shaping the observed spectrum. In particular, the atmospheric gas is not uniformly transparent to radiation across all wavelengths. If the opacity is larger for a given wavelength (e.g., in the core of a spectral absorption line), then the average photon detected at that wavelength originates from higher up in the atmosphere than the average photon at other wavelengths where the opacity is lower.

To a first approximation, the flux emerging at the top of the atmosphere is that of a blackbody spectrum with temperature $T_{\rm eff}$. But in practice, many absorption and scattering processes contribute to shaping a much richer spectrum. Among the main opacity sources there are spectral lines from atomic bound--bound transitions (e.g., hydrogen Balmer lines), bound--free absorption (photoionization), free--free absorption, Thomson scattering from electrons, Rayleigh scattering from atoms and molecules, and molecular opacities (e.g., molecular absorption bands, collision-induced absorption). Precisely characterizing these opacity sources under the conditions found in white dwarf atmospheres remains an active area of research. Of particular interest is the theory of spectral line broadening. In white dwarfs, the width of spectral absorption lines is mostly controlled by collisional broadening (i.e., interactions between the radiating atom and neighbouring particles). 
Describing this accurately is a complex challenge, but the effort is worthwhile as it enables the extraction of valuable information about the physical conditions in the atmosphere from the spectral line shapes \citep{bergeron1992}.

\subsection{Spectral classification}
\begin{figure}
\centering
\includegraphics[width=0.6\textwidth]{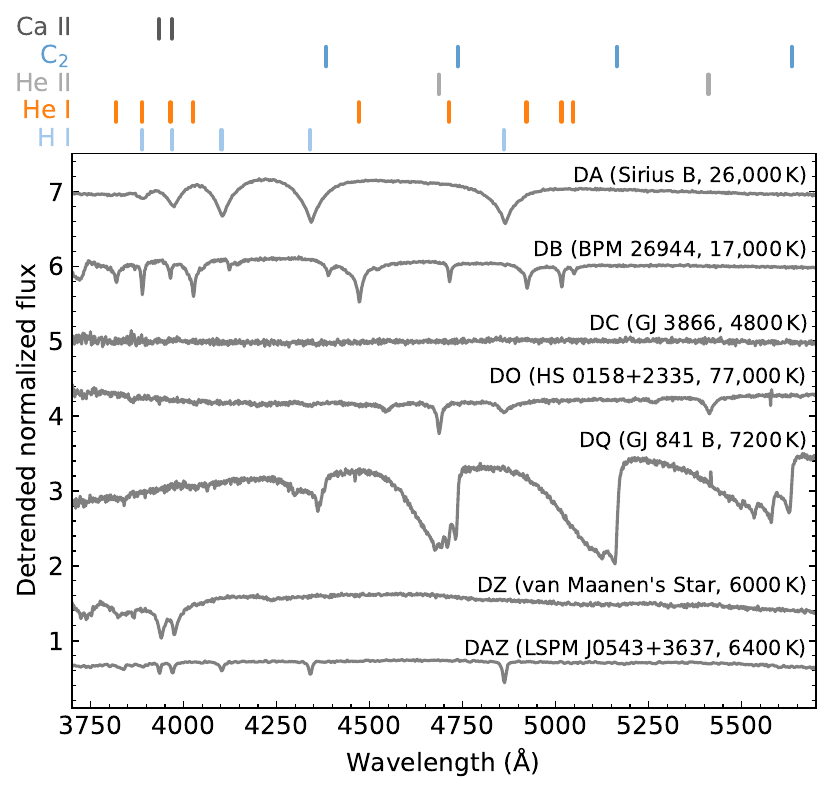}
\caption{Spectra of white dwarfs showcasing diverse spectral types. The spectral type, name of the star and effective temperature are given next to each spectrum. To enhance clarity, linear trends have been subtracted from the original spectra, which are also normalized and vertically offset. The wavelengths of the main atomic absorption lines (or molecular band heads in the case of C$_2$) are indicated above the figure. The spectra shown here are taken from a variety of sources and are freely available on the Montreal White Dwarf Database\protect\footnotemark \citep{dufour2017}.}
\label{fig:sptype}
\end{figure}

The wide range of atmospheric temperatures and compositions found in white dwarfs leads to a diversity of spectral types (Figure~\ref{fig:sptype}). The spectral classification scheme currently used is that of \cite{sion1983}. In its simplest form, the spectral type of a white dwarf is given by the combination of the letter ``D'' (for degenerate) and one of six letters that identify the primary spectral types. The majority of spectroscopically observed white dwarfs belong to the DA spectral class, an example of which is shown at the top of Figure~\ref{fig:sptype}. DA white dwarfs are those that display hydrogen Balmer in their spectra. As we have seen, most white dwarfs have a thick hydrogen layer (Section~\ref{sec:envelope}), and therefore exhibit a pure hydrogen atmosphere, which most of the time produces detectable Balmer lines. Next are the DB white dwarfs, which are characterized by neutral helium spectral lines. DB white dwarfs are much more rare than DAs. This is not only due to the fact that most white dwarfs have hydrogen-dominated atmospheres, but also that helium lines cannot be detected below $T_{\rm eff} \simeq 11{,}000\,$K since there is then too little thermal energy to excite the appropriate electronic transitions. This brings us to the third spectral type, the DC white dwarfs. These are stars that display a featureless, continuous spectrum. Most DC white dwarfs have helium- or hydrogen-dominated atmospheres that are too cool to produce optical absorption lines (the threshold is around 5000\,K for hydrogen). After that, there are the DO white dwarfs, which show ionized helium lines. They are even more rare than DB white dwarfs, as the presence of ionized helium lines requires very hot temperatures where white dwarfs spend very little time due to the initially rapid cooling (Figure~\ref{fig:cooling}).

\footnotetext{\url{https://montrealwhitedwarfdatabase.org/}}

In apparent contradiction to earlier discussions where the outer layers have been described as being composed of hydrogen and/or helium, DQ and DZ white dwarfs are characterized by absorption lines from elements heavier than hydrogen and helium. In the case of DQ white dwarf, that element is carbon, and it can manifests itself either as atomic absorption lines (above $\simeq 10{,}000\,$K) or absorption bands from the C$_2$ molecule (below $\simeq 10{,}000\,$K, such as in the example shown in Figure~\ref{fig:sptype}). The presence of carbon in the atmospheres of most DQ white dwarfs is understood as the result of the convective dredge-up of carbon from the deep interior. This leads to the presence of relatively small traces of carbon in the atmosphere (of the order of one part per million) that are readily detectable thanks to the strong opacity of carbon features. While this dredge-up process successfully explains the presence of carbon in most white dwarfs, it cannot account for the carbon abundances measured in all DQ white dwarfs. Finally, DZ white dwarfs are those that display absorption lines of atomic species other than hydrogen, helium, or carbon. We return to this special case in Section~\ref{sec:pollution}.

White dwarf spectral types can contain more than just two letters. First, additional letters can be appended to the primary spectral class to signal the presence of secondary absorption features. For example, Figure~\ref{fig:sptype} shows the example of a DAZ white dwarf, which has both hydrogen and calcium absorption lines. Second, there are other letters that can be used to signal various peculiarities, such as ``H'' to signal the detection of a magnetic field (more on this in Section~\ref{sec:magnetic}), ``V'' for a variable white dwarf, and ``e'' for emission lines. Before closing this section, it is important to highlight a frequent source of confusion regarding the atmospheric composition and spectral type of a white dwarf. The spectral type is only a description of the appearance of the star's spectrum, not a direct indicator of its atmospheric composition. For example, some DA white dwarfs actually contain more helium than hydrogen in their atmospheres, and most DQs and DZs have helium-dominated atmospheres.

\subsection{Metal pollution}
\label{sec:pollution}
What is responsible for the presence of metals in the atmospheres of DZ white dwarfs (and other similar variants)? We know that the metals frequently observed in the atmospheres of these stars (e.g., Ca, Mg, Fe) should sink out of view below the photosphere in timescales that are very short (days to millions of years) compared to cooling timescales (billions of years). So how do we explain that 25--50\% of white dwarfs show signs of metal pollution \citep{koester2014}? One way to explain this would be the presence of a mechanism that counterbalances gravitational settling. Such a mechanism exists in the form of radiative levitation, but it only operates in very hot stars and can be ruled out for the vast majority of metal-polluted white dwarfs. The alternative explanation, and the correct one, is that these metals arrived in the atmosphere only recently and therefore did not yet have the chance to completely sink below the photosphere. More specifically, metal pollution in white dwarf atmospheres is the result of the accretion of planetary material, an idea that is now very well supported by observations. In fact, some metal-polluted white dwarfs have detectable debris disks \citep{jura2003} or disintegrating planetesimals \citep{vanderburg2015}, and the X-ray signature of the accretion process has now been observed \citep{cunningham2022}.

The spectral analysis of metal-polluted white dwarfs provides a direct window into the chemical composition of the accreted material, offering insights into the makeup of rocky bodies within exoplanetary systems. This is to be contrasted with other methods that infer planetary composition indirectly from bulk density measurement or are limited to surface or atmospheric characteristics. To first order, most metal-polluted white dwarfs show that the accreted material closely resembles the composition of bulk Earth, but there is much more to learn from this unique approach. For instance, variations in the ratios of volatile to refractory elements can provide clues about the conditions under which the accreted bodies were initially formed. Moreover, measuring the composition of bodies accreted by white dwarfs of different ages allows to probe changes in the chemical makeup of rocky bodies throughout the evolution of our galaxy. Needless to say, this field represents an area of very active research, with potential to significantly advance our understanding of the architecture and evolution of planetary systems.

\subsection{Magnetic fields}
\label{sec:magnetic}
In unbiased samples, around 20\% of white dwarfs are found to host a detectable magnetic field \citep{bagnulo2021}. The strength of this field spans orders of magnitude, from a few kilogauss for the weakly magnetic objects all the way to gigagauss levels at the other extreme (for comparison, sunspots have magnetic fields of a few kilogauss). Most of the time, magnetic fields are detected thanks to Zeeman splitting, where the presence of a magnetic field causes the spectral lines of atoms to split into multiple components at different wavelengths. A white dwarf with Zeeman-split Balmer lines is of the DAH spectral type. The magnitude of the Zeeman splitting depends on the strength of the field, thereby allowing a spectroscopic measurement of the field strength. High-resolution spectroscopy may be required to detect very weak fields. Of course, no Zeeman splitting can be observed in featureless DC white dwarfs even if some are strongly magnetic. For these objects, the only option is to use spectropolarimetry, as polarization of the continuum indicates the presence of a magnetic field.

There are striking demographic trends within the magnetic white dwarf population that suggest different origins for magnetic fields in white dwarfs. Notably, more massive white dwarfs exhibit a higher incidence of magnetism compared to their less massive counterparts and have stronger fields. Since a larger fraction of more massive white dwarfs are believed to be the products of stellar mergers, it is possible that the merger process can generate strong magnetic fields. Among normal-mass white dwarfs, it is found that magnetic fields are rarely detected for objects younger than 2--3\,Gyr. However, magnetic white dwarfs become gradually more common at lower temperatures. This progression could be due to the outward diffusion to the surface of a magnetic field generated prior to the white dwarf phase and initially buried in the core. In addition, it is possible that the white dwarf generates a field of its own. This could potentially be achieved during crystallization. When the solid core is formed and has a different composition than the surrounding liquid (Section~\ref{sec:chemevol}), instabilities due to the redistribution of ionic species within the core trigger fluid motions that could generate an internal dynamo \citep{isern2017}. This elegant idea is similar to the mechanism that powers Earth's magnetic field, and testing its validity is an area of active research. 

Finally, it is important to note that magnetic fields have minimal impact on the structure and cooling of white dwarfs. The magnetic pressure remains much smaller than the electron degeneracy pressure in the core unless the internal field strength surpasses $\sim 10^{13}\,$G, which is orders of magnitude stronger than anything observed and hence very unlikely. Therefore, magnetic fields do not affect the mass--radius relation of any known white dwarf. Likewise, magnetic fields smaller than $10^9\,$G are too weak to impact energy transfer at the interface of the degenerate interior (where cooling rates are regulated, see Section~\ref{sec:rhoT}), meaning that these fields cannot significantly alter the cooling process.

\section{Conclusion}
\label{sec:conclusion}
This chapter has endeavored to provide a foundational understanding of white dwarfs. From their discovery and physical nature to the details of their evolutionary history and the diversity of their spectra, we have covered a wide range of topics. However, the scope of this discussion has only scratched the surface of the field. Many exciting topics, some touched upon briefly in this chapter and others barely mentioned, are discussed in more depth elsewhere in this Encyclopedia. This includes the initial--final mass relation that prescribes the masses of white dwarfs, the burgeoning field of white dwarf asteroseismology, and white dwarf exoplanetary systems. Most egregiously, we have not discussed the vast and rich topic of white dwarfs in binary systems, which encompasses a broad spectrum of phenomena from Type Ia supernovae to cataclysmic variables and gravitational waves. It is the author's hope that this chapter has furnished the reader with the essential background required to delve deeper into the many fascinating aspects of white dwarfs.


\begin{ack}[Acknowledgments]

The author thanks the Canadian Institute for Theoretical Astrophysics (CITA) National Fellowship program for financial support.
\end{ack}


\bibliographystyle{Harvard}
\bibliography{reference}

\section*{Further information}
\begin{itemize}
    \item For a classic (although now somewhat dated) review article on white dwarfs: \cite{fontaine2001};
    \item For an accessible historical account of the development of the field of white dwarf studies: \cite{vanhorn2015};
    \item For a recent review focusing on the physics of white dwarfs: \cite{saumon2022}.
\end{itemize}

\end{document}